\documentclass[useAMS,usenatbib]{mn2e}

\usepackage{graphicx}
\usepackage{txfonts}
\usepackage{times,epsfig} 
\usepackage[dvips]{color}
\usepackage[]{natbib}
\usepackage{latexsym}
\usepackage[italian,english]{babel}
\usepackage[latin1]{inputenc}
\usepackage{epstopdf}

\usepackage[fleqn]{amsmath}
\usepackage{amsfonts}
\usepackage{amssymb}
\usepackage{amstext}

\usepackage{amsthm}
\usepackage{bm}
\usepackage{amsopn}
\usepackage{xspace}
\usepackage{array}
\usepackage{epsfig}
\usepackage{rotating}
\usepackage{subfig}
\usepackage{url}
\usepackage[T1]{fontenc}
\usepackage{lmodern}
 
\usepackage{tabularx}
\usepackage{longtable}
\usepackage{textcomp}
\usepackage{pdflscape}
\usepackage{rotating}
\usepackage{setspace}
\usepackage{footmisc}
\usepackage{booktabs}
\usepackage{multirow}
\usepackage{booktabs}
\usepackage{tablefootnote}
\usepackage[section]{placeins}

\title[Do radio mini-halos and gas heating in cool-core clusters have a common origin?]{Do radio mini-halos and gas heating in cool-core clusters have a common origin?}
\author[]{L. Bravi$^{1,2}$, M. Gitti$^{1,2}$, G. Brunetti$^{2}$\thanks{E-mail:luca.bravi@studio.unibo.it} 
 \\ 
$^{1}$Physics and Astronomy Department, University of Bologna, via Ranzani 1, 40127 Bologna, Italy\\
$^{2}$INAF, Osservatorio di Radioastronomia, via Gobetti 101, I-40129 Bologna, Italy}

\begin{document}

\date{}

\pagerange{\pageref{firstpage}--\pageref{lastpage}} \pubyear{2015}

\maketitle

\label{firstpage}

\vspace{-0.15cm}
\begin{abstract}
\vspace{-0.01in}

\noindent
In this letter we present a study of the central regions of cool-core
clusters hosting radio mini-halos, which are diffuse synchrotron
sources extended on cluster-scales surrounding the radio-loud
brightest cluster galaxy. We aim to investigate the interplay
between the thermal and non-thermal components in the intra-cluster
medium in order to get more insights into these radio sources, whose
nature is still unclear. It has recently been proposed that
turbulence plays a role for heating the gas in cool cores. By
assuming that mini-halos are powered by the same turbulence,
we expect that the integrated radio luminosity of mini-halos, $\nu
P_{\nu}$, depends on the cooling flow power, $P_{\rm CF}$, which in
turn constrains the energy available for the non-thermal components
and emission in the cool-core region. We carried out a homogeneous
re-analysis of X-ray {\it Chandra} data of the largest sample of
cool-core clusters hosting radio mini-halos currently available
($\sim$ 20 objects), finding a quasi-linear correlation, $\nu P_{\nu}
\propto P_{\rm CF}^{0.8}$. We show that the scenario of a common
origin of radio mini-halos and gas heating in cool-core clusters is
energetically viable, provided that mini-halos trace regions where the
magnetic field strength is $B \gg 0.5\, \mu$G .
\end{abstract}



\vspace{-0.15in}
\section{Introduction}
\vspace{-0.01in}

The amount of gas in cool-core clusters that is cooling radiatively to
low temperatures is found to be much less
%
%
than what is predicted by the standard cooling flow model
\citep[e.g.,][for reviews]{Fabian_1994,Peterson_2006}. The
implication is that the central intra-cluster medium (ICM) of these
``cool-core clusters'' must experience some kind of heating to balance
cooling. The most promising source of heating has been identified as
feedback from energy injection by the active galactic nucleus (AGN) of
the brightest cluster galaxy (BCG) \citep[e.g.,][and reference
therein]{Mcnamara_2007,Mcnamara_2012,Gitti_2012,Fabian_2012}. At the
same time, mechanically-powerful AGN are likely to drive turbulence in
the central ICM which may contribute to gas heating. In this context,
\citet{Zhuravleva_2014} recently found that the AGN-driven turbulence
must eventually dissipate into heat and it is sufficient to offset
radiative cooling. On the other hand, such turbulence can also play
a role for particle acceleration and magnetic field amplification in
the ICM.

Diffuse synchrotron emission has been observed in a number of
cool-core clusters in the form of ``radio mini-halos'' surrounding the
radio-loud BCG \citep[e.g.,][for reviews]{Feretti_2012,
  Brunetti_2014}. Mini-halos, which have steep ($\alpha \sim
1.1$, $S(\nu) \propto \nu^{-\alpha}$) radio spectra and
amorphous (roundish) shape, extend on scales $\sim 100-500$ kpc (total
size) tracing regions where the ICM cooling time is short and the ICM
is compressed.
The origin of mini-halos is still unclear and it has generated a
lively discussion in the last decade \citep[e.g.,][]{Brunetti_2014}. One possibility is that they form through the
re-acceleration of relativistic particles by turbulence
\citep[][]{Gitti_2002,Gitti_2004,Mazzotta_2008,Zuhone_2013}. Alternatively
  they may be of hadronic origin \citep[][]{Pfrommer_2004, Zandanel_2014}.

In this letter we assume a re-acceleration scenario where the
turbulence is responsible for both the origin of mini-halos and for
quenching cooling flows. In the framework of this scenario, we expect
a connection between the cooling flow power, $P_{\rm CF}$, and the
mini-halo integrated radio power, $\nu P_{\nu}$. A trend between $\nu
P_{\nu}$ and $P_{\rm CF}$ was observed by
\citet{Gitti_2004,Gitti_2012} using small, heterogeneous samples of
mini-halos. On the other hand, in recent years mini-halos are being
found in an increasing number of cool-core clusters
\citep[e.g.,][]{Govoni_2009, Giacintucci_2014}, thus allowing a
substantial step in the field. The aim of this work is to overcome the
limitations in the previous studies by exploiting the increased sample
statistics in order to obtain more insights into the origin of
mini-halos. In particular, we present the results of a homogeneous
re-analysis of \textit{Chandra} data of the largest collection of
mini-halo clusters currently known ($\sim$ 20 objects,
Sect.~\ref{sec:sample}), and investigate the connection between the
thermal properties of cool cores and the non-thermal properties of
mini-halos (Sect.~\ref{sec:results}). We further discuss the
consistency of the adopted turbulent model and derive constraints on
the magnetic field in the mini-halo region (Sect.~\ref{sec:result}).
We adopt a $\mathrm{\Lambda CDM}$ cosmology with $\mathrm{H_{0}=70}$
km $\mathrm{s^{-1}Mpc^{-1}}$, $\Omega_{M} = 1 - \Omega_{\Lambda} =
0.3$.

\vspace{-0.15in}
\section{Mini-Halo sample and X-ray data}
\label{sec:sample}

\subsection{Sample selection}
\vspace{-0.01in}

Our sample is obtained from the list of 21 mini-halos reported in
\citet{Giacintucci_2014}, who recently selected a large collection of
X-ray-luminous clusters from the {\it Chandra} ACCEPT\footnote{Archive
  of Chandra Cluster Entropy Profiles Tables.} sample
\citep[]{Cavagnolo_2009} with available high-quality radio data from
archival VLA (Very Large Array) and GMRT (Giant Metrewave Radio
Telescope) observations, and discovered four new mini-halos. We
further included the new mini-halo detection in the Phoenix cluster
\citep{vanWeeren_2014}.

X-ray {\it Chandra} archival observations are available for all
clusters. To ensure a uniform quality to the X-ray data, we excluded shallow observations,
with an exposure time $<$ 20 ks. Furthermore, we have considered only
the observations where the cluster core emission, which corresponds to
the location of the cooling region and of the mini-halo we are
interested in, is well pointed and enclosed in the central chip. This
guarantees that the data reduction process is performed in a
homogeneous, consistent manner for all the objects in our sample. The
full mini-halo sample used in this work finally comprises 20 objects,
listed in Table~\ref{tab:radio_MH}, where we report the radio
properties taken from the literature
\citep[][]{Giacintucci_2014,vanWeeren_2014}. By excluding the objects
classified by \citet[][]{Giacintucci_2014} as ``candidate'' or
``uncertain'', we further selected a sub-sample of 16 confirmed
mini-halos.

\vspace{-0.15in}
\subsection{Chandra data preparation}
\vspace{-0.01in}

Data were reprocessed with CIAO 4.6, using CALDB 4.5.9 and corrected
for know time-dependent gain problems following techniques similar to
those described in the \textit{Chandra} analysis
threads\footnote{http://cxc.harvard.edu/ciao/threads/index.html.}. Screening
of the event files was applied to filter out strong background
flares. Blank-sky background files, filtered in the same manner as in
each cluster and normalized to the count rate of the source image in
the 9.0-12.0 keV band, were used for background subtraction. We
identified and removed the point sources in the CCD using the CIAO
task \texttt{WAVEDETECT}. Images, instrument maps, and exposure maps
were created in the 0.5 - 7.0 keV band. Data with energies above 7.0
keV and below 0.5 keV were excluded in order to prevent background
contamination and uncertainties in the ACIS calibration, respectively.

\vspace{-0.15in}
\section{Spectral analysis and results}
\label{sec:results}

\subsection{Cool-Core spectral analysis}
\label{sec:core}
\vspace{-0.01in}

In order to extract the azimuthally averaged profiles of the physical
parameters of the thermal ICM, we created concentric annuli centred on
the peak of the X-ray emission of each cluster. For each annulus was
extracted a single spectrum that was then modelled using the XSPEC
code, version 12.8.1g. Spectral fitting was performed in the [0.5 - 7]
keV band. In order to correct the projection effects we fitted the
spectra using a \texttt{projct*wabs*apec} model. The free parameters
in this model are temperature \textit{kT}, metallicity \textit{Z}
(measured relative to the solar values) and normalization parameters
of the \texttt{apec} model. The hydrogen column density was fixed at
the Galactic value \citep[]{Dickey_1990}. The deprojected fits allow us to derive a radial
profile of the temperature, \textit{kT}, and of the electron density,
$n_{\rm e}$. With this two quantities, we estimated the cooling time
as the time necessary for the ICM to radiate its enthalpy per unit
volume:
\begin{eqnarray}
t_{{\rm cool}} = \dfrac{5}{2}\dfrac{kT}{\mu X_{H}n_{e}\Lambda (T)}\label{eq:T_cool}
\end{eqnarray}
where $\mu = 0.61$ is the molecular weight for a fully ionized plasma,
$X_{H} = 0.71$ is the hydrogen mass function, and $\Lambda (T)$ is the
cooling function (we have interpolated the table by
\citet{Sutherland_1993} as a function of temperature and metallicity
Z). The cooling radius $r_{{\rm cool}}$ is traditionally defined as
the radius at which $t_{{\rm cool}}$ is equal to the age of the
systems, usually taken to be the look-back-time at $\textit{z}=1$,
$t_{{\rm cool}} \sim 7.7$ Gyr. However, in this work we adopted a
shorter time interval in which the system has realistically been
relaxed, i.e. the time since the last merger event, $t_{{\rm cool}} =
3$ Gyr. Accounting for the different definitions of $t_{{\rm cool}}$,
our estimates, reported in Table~\ref{tab:radio_MH}, are in agreement
with the cooling radii of the ACCEPT sample
\citep[]{Cavagnolo_2009}. We note that the radius of the mini-halo,
$R_{\rm MH}$, is generally larger than our definition of $r_{{\rm cool}}$,
corresponding in cooling times in the range 4 Gyr $\div$
$t_{\rm Hubble}$. This readly implies that the region of strong cooling is
smaller than the mini-halo extension.

\vspace{-0.15in}
\subsection{Spectral analysis in the mini-halo region}
\label{sec:core}
\vspace{-0.01in}

In order to determine the physical properties of the thermal ICM in
the region where the diffuse radio emission is present, we extracted a
single spectrum inside $R_{{\rm MH}}$ for each cluster of the
sample. The spectra are modelled using a \texttt{wabs*(apec+mkcflow)}
model\footnote{We also used a \texttt{projct*wabs*(apec+mkcflow)}
  model that corrects for the contribution of the foreground emission
  projected along the line-of-sight. The results are consistent within
  the errors with those of the projected model
  \texttt{wabs*(apec+mkcflow)}.}. The model assumes a combination of a
standard single temperature emission (\texttt{apec} component) and of
a multi-phase component that takes into account a isobaric cooling
flow emission, \texttt{mkcflow} \citep[]{Arnaud_1996}. This model fit
provides a direct estimate of the amount of gas that is cooling. Under
the assumptions that the thermal component represents the ambient
cluster atmosphere and that the cooling flow component is cooled
ambient gas, the higher temperature and metallicity parameters of the
\texttt{mkcflow} component were tied to those of the \texttt{apec}
component. Contrary to the previous spectral analysis in concentric
annuli, here the hydrogen column density was not fixed to the Galactic
value since a different best-fit value was often preferred by the
fit. Furthermore, according to the physics of a standard cooling flow
model, the lower temperature was fixed to the lowest possible value
($\sim 0.1$ keV). The (free) normalization parameter of the
\texttt{mkcflow} model is the mass deposition rate $\dot{M}$. The
values of $\dot{M}$ that we obtained are in line with typical values
from the literature and with independent estimates derived for some of
our clusters by different authors \citep[e.g.,][]{Rafferty_2006}.
The best-fitting parameter values and the 90\% confidence
level derived for each cluster are summarized in
Table~\ref{tab:radio_MH}.
\begin{table*}
\caption{Properties of our sample of mini-halo clusters.}
\label{tab:radio_MH}      
{\centering
\begin{tabular}{lccccccccc}
\hline
\midrule
			Cluster name & \textit{z} &  $S_{{\rm MH \, [1.4GHz]}}$ & $r_{{\rm cool}}$ & $R_{{\rm MH}}$ & $\nu P_{\nu \, {\rm [1.4GHz]}}$ & \textit{kT} & $\dot{M}$ & $P_{\rm CF}$ & Notes\\
 			 & & (mJy) & (kpc) & (kpc) & $\mathrm{10^{40} \; erg \, s^{-1}}$ & (keV) & ($M_{\odot} yr^{-1}$) & $10^{44} \, \mathrm{erg \, s^{-1}}$ &\\
			\midrule
			2A 0335+096 & 0.035 & $21.1\pm 2.1$ & $46\pm 1$ & 70 & $0.08\pm 0.01$ & $2.85^{+0.02}_{-0.03}$ & $111.9^{+4.8}_{-4.7}$ & $ 0.32\pm 0.02$ &\\[0.4ex]
			A 2626 & 0.055 & $18.0\pm 1.8$ & $17\pm 2$ & 30 & $0.19\pm 0.01$ & $2.95^{+0.24}_{-0.23}$ & $4.3^{+1.8}_{-1.7}$ & $0.02\pm 0.01$ &U\\[0.4ex]
			A 1795 & 0.063 & $85.0\pm 4.9$ & $39\pm 2$ & 100 & $1.11\pm 0.07$ & $4.62^{+0.08}_{-0.04}$ & $21.1^{+6.7}_{-8.8}$ & $0.10\pm 0.04$ &C\\[0.4ex]
			ZwCl 1742.1+3306 & 0.076 & $13.8\pm 0.8$ & $32\pm 1$ & 40 & $0.28\pm 0.01$ & $3.04^{+0.11}_{-0.10}$ & $30.1^{+5.7}_{-5.4}$ & $0.09\pm 0.02$& U\\[0.4ex]
			A 2029 & 0.077 & $19.5\pm 2.5$ & $35\pm 1$ & 270 & $0.39\pm 0.06$ & $7.58^{+0.11}_{-0.11}$ & $9.6^{+8.9}_{-6.0}$ & $0.07\pm 0.05$ &\\[0.4ex]
			A 478 & 0.088 & $16.6\pm 3.0$ & $45\pm 1$ & 160 & $0.45\pm0.08$ & $6.01^{+0.14}_{-0.14}$ & $20.8^{+36.8}_{-20.8}$ & $ < 0.34$ &\\[0.4ex]
			A 2204 & 0.152 & $8.6\pm 0.9$ & $50\pm 1$ & 50 & $0.76\pm 0.07$ & $4.21^{+0.08}_{-0.08}$ & $< 0.01$ & $< 0.06$ &\\[0.4ex]
			RX J1720.1+2638 & 0.159 & $72.0\pm 4.4$ & $46\pm 2$ & 140 & $7.46\pm 0.45$ & $5.05^{+0.18}_{-0.17}$ & $0.02^{+38.80}_{-0.01}$ & $< 0.20$ &\\[0.4ex]
			RXC J1504.1-0248 & 0.215 & $20.0\pm 1.0$ &
                        $64\pm 1$ & 140 & $3.78\pm 0.20$ &
                        $5.87^{+0.35}_{-0.30}$ &
                        $606.2^{+419.5}_{-396.0}$ & {\bf $3.5\pm 2.4$} &\\[0.4ex]
			A 2390 & 0.228 & $28.3\pm 4.3$ & $38\pm 1$ & 250 & $6.24\pm 0.94$ & $8.01^{+0.39}_{-0.36}$ & $69.0^{+98.5}_{-68.0}$ & $< 1.34$ &\\[0.4ex]
			A 1835 & 0.252 & $6.1\pm 1.3$ & $57\pm 1$ & 240 & $1.66\pm 0.35$ & $7.64^{+0.35}_{-0.14}$ & $295.2^{+154.5}_{-81.3}$ & $2.25\pm 0.90$ &\\[0.4ex]
			MS 1455.0+2232 & 0.258 & $8.5\pm 1.1$ & $55\pm 1$ & 120 & $2.45\pm 0.32$ & $4.61^{+0.26}_{-0.23}$ & $242.4^{+206.1}_{-187.5}$ & $1.11\pm 0.91$ &\\[0.4ex]
			ZwCl 3146 & 0.290 & $\sim 5.2$ & $60\pm 1$ & 90 & $1.95\pm0.01$ & $5.73^{+0.37}_{-0.29}$ & $276.7^{+165.4}_{-149.3}$ & $1.58\pm 0.90$ &\\[0.4ex]
			RX J1532.9+3021 & 0.345 & $7.5\pm 0.4$ & $67\pm 1$ & 100 & $4.69\pm 0.24$ & $5.28^{+0.57}_{-0.30}$ & $647.8^{+411.8}_{-331.1}$ & $3.41\pm 1.98$ &\\[0.4ex]
			MACS J1931.8-2634 & 0.352 & $47.9\pm 2.8$ & $61\pm 1$ & 100 & $28.0\pm 1.7$ & $6.11^{+1.01}_{-0.60}$ & $1178.0^{+524.9}_{-511.0}$ & $7.17\pm 3.30$ &U\\[0.4ex]
			RBS 797 & 0.354 & $5.2\pm 0.6 $ & $69\pm 1$ & 120 & $3.08\pm 0.34$ & $5.46^{+0.42}_{-0.28}$ & $59.2^{+491.9}_{-59.1}$ & $< 3.00$ &\\[0.4ex]
			MACS J0159.8-0849 & 0.405 & $2.4\pm 0.2$ & $53\pm 1$ & 90 & $1.95\pm 0.20$ & $6.59^{+3.32}_{-0.79}$ & $106.7^{+484.3}_{-106.7}$ & $< 3.90$ & C\\[0.4ex]
			MACS J0329.6-0211 & 0.450 & $3.8\pm 0.4$ & $54\pm 1$ & 70 & $3.98\pm 0.42$ & $4.83^{+1.13}_{-0.52}$ & $126.6^{+569.3}_{-126.6}$ & $< 3.36$ & C\\[0.4ex]
			RX J1347.5-1145 & 0.451 & $34.1\pm 2.3$ & $62\pm 2$ & 320 & $35.8\pm 2.5$ & $21.6^{+14.3}_{-4.6}$ & $2852.3^{+399.1}_{-473.7}$ & $61.3\pm 29.1$ &\\[0.4ex]
			Phoenix & 0.596 & $6.8\pm 2.0$ & $73\pm 1$ & 176 & $14.1\pm 4.2$ & $10.8^{+14.5}_{-2.6}$ & $4353.2^{+3744.0}_{-4353.0}$ & $< 121.81$ &\\[0.4ex]
			\midrule
			\multicolumn{10}{l}{%
  \begin{minipage}{0.98\textwidth}
    \footnotesize \textbf{Notes:} Col. (1): Cluster name. Col. (2):
    Redshift. Col. (3): Mini-halo flux density at 1.4 GHz from
    \citet{Giacintucci_2014}, except in the case of Phoenix where the
    value was estimated from the observations at 610 MHz
    \citep{vanWeeren_2014} by assuming a spectral index of $\alpha =
    1.1$. Col. (4): Cooling radius corresponding to a cooling time of
    3 Gyr. Col. (5): Average radius of the mini-halo estimated by
    \citet{Giacintucci_2014} as $R_{\rm MH} = \sqrt{R_{{\rm max}}
      \cdot R_{{\rm min}}}$, where $R_{\rm max}$ and $R_{\rm min}$ are
    the maximum and minimum radius as derived from the $+3\sigma$
    isocontour of the image. For consistency, we have used this
    equation to estimate $R_{\rm MH}$ of the Phoenix cluster from the
    published maps of \citet{vanWeeren_2014}. Col. (6): Radio power of
    mini-halos at 1.4 GHz (in terms of integrated radio luminosity,
    $\nu P_{\nu}$). Col. (7): Temperature of the \texttt{apec}
    component of the \texttt{wabs*(apec+mkcflow)} spectral model
    inside $R_{\rm MH}$. Col. (8): Mass accretion rate derived from
    the normalization parameter of the \texttt{mkcflow}
    component inside $R_{\rm MH}$. Col. (9): Cooling flow power estimated as $P_{{\rm
        CF}} = \frac{\dot{M}kT}{\mu m_{p}}$ inside $R_{\rm MH}$. Col. (10): Uncertain (U)
    and candidate (C) mini-halos according to \citet{Giacintucci_2014}.
  \end{minipage}}
\\
		\end{tabular}}
\end{table*}

\vspace{-0.15in}
\subsection{The correlation between $\nu P_{\nu}$ and $P_{{\rm CF}}$}
\label{sec:cfp}
\vspace{-0.01in}

\citet{Gitti_2004,Gitti_2012} found a correlation between the radio
power of mini-halos at 1.4 GHz (in terms of integrated radio
luminosity, $\nu P_{\nu}$), and the cooling flow power, $P_{{\rm
    CF}}$. The maximum power $P_{\rm CF}$ available in the cooling
flow can be estimated assuming a standard cooling flow model and it
corresponds to the $p dV/dt$ work done on the gas per unit time as it
enters $r_{\rm cool}$:
\begin{equation}
P_{\rm CF} =  \dfrac{\dot{M} kT}{\mu m_{{\rm p}}} 
\label{eq:Pcf_lx}
\end{equation}
\citep[e.g.,][]{Fabian_1994,Gitti_2004}, where $\dot{M}$ is the mass
deposition rate and $kT$ is the
temperature of the gas at $r_{\rm cool}$. 
%
\begin{figure}
\centering
\includegraphics[width=8.6cm]{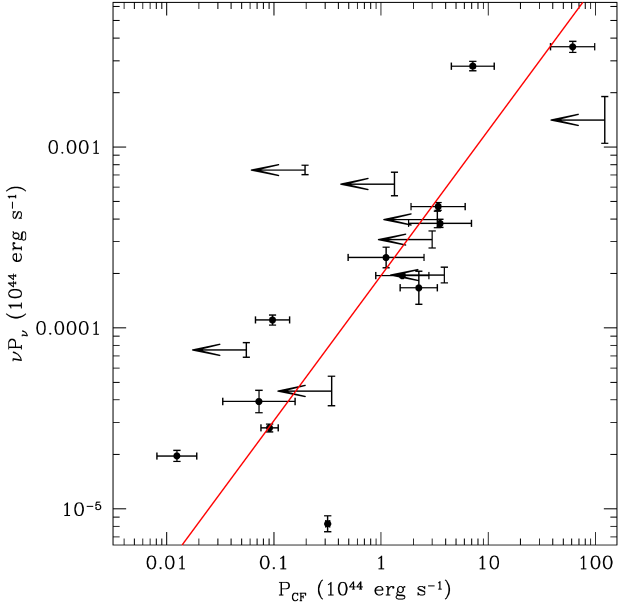}
\caption{Correlation between the radio emitted power of mini-halos at
  1.4 GHz, in terms of $\nu P_{\nu}$, and the cooling flow power,
  $P_{{\rm CF}}$, for our sample of mini-halo clusters (see columns 6
  and 9 in Table~\ref{tab:radio_MH}). The black arrows show the upper limits when $P_{{\rm CF}}$ is not
    constrained. The red line represents the best-fit relation determined without the upper limits on $P_{{\rm
        CF}}$.}\label{fig:pcf}
\end{figure}
However, to compare powers emitted inside the same volume, i.e. that
of the mini-halo, in this work we estimated $P_{\rm CF}$ inside
$R_{{\rm MH}}$. In particular, we estimated $\dot{M}$ from the
normalization of the \texttt{mkcflow} component and $kT$ from the
temperature of the \texttt{apec} component\footnote{The temperature
  $kT$ used here is in agreement with that estimated at $R_{\rm MH}$
  from the radial temperature profile (Sec.~\ref{sec:core}).} derived
in the previous section. The values of $P_{{\rm CF}}$ are reported in
Table~\ref{tab:radio_MH}.

The correlation between the radio emitted power of mini-halos at 1.4
GHz, in terms of $\nu P_{\nu}$, and the cooling flow power, $P_{{\rm
    CF}}$, is shown in Fig.~\ref{fig:pcf}.  
We used the bivariate correlated error and intrinsic scatter (BCES)
algorithm \citep[]{Akritas_1996} to perform regression fits in log
space to the data of the 12 clusters for which $P_{{\rm CF}}$
  is constrained, determining the best-fitting powerlaw relationship
(bisector method)\footnote{We note that the upper limits on
    $P_{{\rm CF}}$ estimated for the other clusters are not in
    disagreement with the correlation.}:
\begin{equation}
\log(\nu P_{\nu})=[(0.80\pm 0.13) \cdot \, \log(P_{\rm CF})]-(3.70\pm 0.11)
\label{eq:fit1}
\end{equation}
We used a Spearman test to evaluate the strength of the
  correlation. The Spearman parameters are $r_{s}$ and
  $\textit{probrs}$, where the former is a non-parametric measure of
  the statistical dependence between two variables and the latter
  is the two-sided significance level of deviation from zero. High
  values of $r_{s}$ and small values of $\textit{probrs}$ indicate a
  significant correlation. For our data the Spearman test parameters
are: $r_{s} = 0.89$ and $\textit{probrs}=1.14 \times 10^{-4}$, thus
confirming the strength of the correlation.
\\
%
The correlation is present also in the sub-sample of confirmed
mini-halos\footnote{The sub-sample of confirmed mini-halos
    includes all clusters except those labeled with U and C in Table~\ref{tab:radio_MH}.} (Spearman test values: $r_{s} =
0.86$, $\textit{probrs}=6.5 \times 10^{-3}$):
\begin{equation}
\log(\nu P_{\nu})=[(0.76\pm 0.15) \cdot \, \log(P_{\rm CF,})]-(2.12\pm 0.56)
\label{eq:fit3}
\end{equation}
and it is in agreement with the best-fit relations of the full sample.

\begin{figure}[t]
\includegraphics[width=8.6cm]{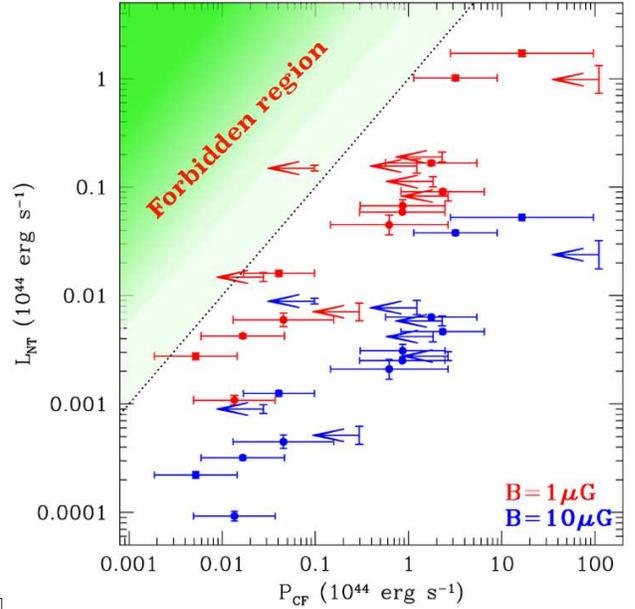}
\caption{Correlation between $L_{\rm NT}$ and $P_{\rm CF}$ for our
  mini-halo cluster sample. The red and blue points were calculated
  assuming $B =1\, \mu$G and $B =10\, \mu$G, respectively. The black
  dotted line represents the 1:1 relation. The upper side of 1:1 relation is the forbidden region, where $L_{\rm NT} > P_{\rm H} \sim P_{\rm CF}$.}\label{fig:cons_B}
\end{figure}

As final sanity checks, we verified that the correlation exists also
if $P_{{\rm CF}}$ is estimated as $P_{\rm CF} = 2/5 \cdot
L_{\rm cool}$ \citep[e.g.,][]{Fabian_1994,Gitti_2004}, where $L_{\rm
  cool}$ is the cooling X--ray luminosity estimated as the total
bolometric luminosity of the \texttt{wabs*(apec+mkcflow)} spectral
model fitted inside $R_{\rm MH}$. Furthermore, to ensure that
  the trend observed in the radio luminosity - X--ray luminosity plane
  is not biased by any effects due to the cluster redshift, we checked
  that a correlation is present also in the radio flux - X--ray flux
  plane for the 12 clusters with well-constrained values. Finally,
all correlations mentioned above are present also if $P_{{\rm CF}}$ is
derived from the spectral analysis inside $r_{{\rm cool}}$.

\vspace{-0.15in}
\section{Discussion}
\label{sec:result}
\vspace{-0.01in}

Starting from our mini-halo sample, having taken as robust an
  approach to the data as possible, we found that there are currently
  only a dozen clusters for which $P_{\rm CF}$ can be constrained from
  XSPEC spectral fitting (see Table~\ref{tab:radio_MH}). Using these
  clusters we have confirmed a correlation between $\nu P_{\nu}$ and
$P_{\rm CF}$, revealing a connection between the energy-reservoir in
cooling flows and that associated to the non-thermal components
powering radio mini-halos.

A solution proposed for the cooling flow problem is based on a
mechanism of heating that is distributed in the core (that is
comparable to the size of radio mini-halos). This mechanism must be
gentle, dissipating energy in the form of heat at a rate, $P_{\rm H}$,
that cannot be much larger than the cooling power, otherwise cool
cores would be disrupted. This is $P_{\rm H} \gtrsim P_{\rm CF}$.
Various sources of energy capable of compensating cooling have been
suggested, the most promising being heating by the feedback due to AGN
\citep[e.g.,][for reviews]{Mcnamara_2007, Fabian_2012, Gitti_2012},
although details are still unclear. Recently \citet{Zhuravleva_2014}
analysed deep X-ray {\it Chandra} data of the Perseus and Virgo
clusters and found that heating by turbulent dissipation evaluated in
the ICM appears to balance radiative cooling locally at each
radius. They suggested that turbulent dissipation may be the key
mechanism responsible for compensating gas cooling losses and keeping
cluster cores in an approximate steady state.

Turbulence is also proposed as an important player for the origin of
mini-halos \citep[leptonic models,][]{Gitti_2002, Mazzotta_2008,
  Zuhone_2013}, although the origin of the turbulence and its
  connection with the thermal and dynamical properties of cool-core is
  still unclear.  Here we argue that particle acceleration and gas
heating in cool-cores are due to the dissipation of the same
turbulence. Obviously this is a simplified picture. Indeed several
turbulent components are probably generated in cool cores and may
contribute in different ways to the heating of the gas and to the
reacceleration of relativistic particles (see Brunetti \& Jones 2014
for a review in the ICM).  Anyhow, if we assume this simple picture, a
fraction of $P_{\rm H}$ will be channelled into particle acceleration
and non-thermal radiation. In this case, assuming $P_{\rm H} \gtrsim
P_{\rm CF}$, Eq. \ref{eq:Pcf_lx} provides an upper limit to the
non-thermal radiation, $L_{\rm NT}$, that can be maintained in the
region of radio mini-halos for a time-scale that is longer than the
radiative life-time of the relativistic electrons.

\noindent
The non-thermal radiation is :
\begin{eqnarray}
L_{\rm NT} = L_{\rm Syn} + L_{\rm IC} = L_{\rm Syn}\left[1+\left(\dfrac{B_{\rm CMB}}{B}\right)^2\right]\label{eq:lum_non_term}
\end{eqnarray}
where $L_{\rm Syn}$ is the synchrotron luminosity, $L_{\rm
    IC}$ is the inverse Compton luminosity, $B_{\rm CMB} = 3.2 \,
(1+z)^2 \, \mu$G is the magnetic field equivalent to the inverse
Compton losses with CMB photons and $B$ is the magnetic field
intensity in the mini-halo region.

In Fig. \ref{fig:cons_B} we report $L_{\rm NT}$ of mini-halos in our
sample, estimated by assuming two reference values of $B=1\, \mu$G and
$B=10\, \mu$G, versus $P_{\rm CF}$ measured in the hosting clusters.
Magnetic fields of the order of 10 $\mu$G are estimated in cool-core
clusters from current Faraday rotation studies \citep[][for
reviews]{Carilli-Taylor_2002, Feretti_2012}. In this case the scenario
is found energetically consistent, namely mini-halos remain distant
from the forbidden region, where $L_{\rm NT} > P_{\rm H} \sim P_{\rm
  CF}$.  On the other hand, for weak magnetic fields, $B < 0.5\, \mu$G,
we find that $L_{\rm NT} \gtrsim P_{\rm CF}$ implying that the
scenario becomes not plausible. Obviously this limit can be released
if we assume more complex situations where multiple turbulent
components coexist in cool cores and that reacceleration of
relativistic particles and gas heating are powered by different
components.  On the other hand, we note at the same time that
relativistic particles get in general only a small fraction of the
turbulent energy flux \citep[see
however][]{Brunetti-Lazarian_2011}. Consequently in the presence of
weak fields, we should admit an unlikely situation where an energetic
turbulent component, with specific turbulent energy $\epsilon_{\rm t}
> P_{\rm CF}/M_{\rm gas}$,
coexists with the gas without disturbing gas thermodynamics.

However, weak magnetic fields do not challenge only reacceleration
models.  Also alternative models, such as the hadronic models
\citep[e.g.,][]{Pfrommer_2004}, are challenged in the case of $B <$
few $\mu$G, due to current gamma-ray upper limits to the $\pi^o$-decay
emission obtained for nearby clusters hosting mini-halos
\citep[e.g.,][]{Aleksic_2012}.  Consequently new theoretical scenarios
will be needed if future observations provide evidence for weak
magnetic fields in cool-core clusters hosting diffuse radio emission.
Future observations with ASTRO-H in the hard X-rays and Faraday
rotation studies with the new radio facilities, such as the JVLA and
SKA precursors, are expected to contribute to constrain magnetic
fields in cool-core clusters.  In particular several Faraday
  rotation measure survey experiments are already planned such as the
  POSSUM survey on ASKAP \citep[][]{Gaensler_2010}, the JVLA
  polarization survey, VLASS \citep[][]{Myers_2014} and the
  forthcoming all-sky polarimetric survey and associated Rotation
  Measure grid to be carried out on SKA1\_MID
  \citep[][]{Johnston_2015}.

\vspace{-0.15in}
\section{Summary and Conclusions}
\vspace{-0.01in}

In this work we have overcome the limitations in the previous studies
by exploiting the increased statistics of known radio mini-halos that
allows us to obtain further insights on their origin. In particular, 
we have carried out an homogeneous analysis of archival
  \textit{Chandra} data of the largest existing sample of mini-halo
  clusters (20 objects) in order to study the X-ray properties of cool
  cores hosting radio mini-halos. Our
main results can be summarized as follows: 

\vspace{-0.05in}
\begin{itemize}

\item We estimated the cooling flow power, $P_{\rm CF}$, inside the
  mini-halo region, and compared it with the radio emitted power of
  mini-halos at 1.4 GHz, in terms of $\nu P_{\nu}$. By using
    the 12 clusters for which the value of $P_{\rm CF}$ is
    constrained, we found a correlation $\nu P_{\nu} \propto P_{\rm
    CF}^{0.8}$. This suggests a connection between the thermal
  properties of cool-core clusters and the non-thermal properties of
  mini-halos, confirming the previous results obtained by
  \citet{Gitti_2004,Gitti_2012}.

\item We discussed a scenario where turbulence in the cool cores is
  responsible for both the origin of mini-halos and for the solution
  of the cooling flow problem. In this context, $P_{\rm CF}$ can be
  regarded as an upper limit to the non-thermal luminosity $L_{\rm
    NT}$ generated in the mini-halo region.  The limit $P_{\rm CF}
  \gg L_{\rm NT}$ allows us to set a corresponding lower limit $B >
  0.5\,  \mu$G to the typical magnetic field in mini-halos.

\end{itemize}
\vspace{-0.05in}

Future efforts and observations with ASTRO-H, JVLA, SKA-pathfinders
and precursor are essential to build large mini-halo samples and
achieve a full understanding of the mechanism for the origin of
mini-halos \citep[see e.g.,][for a recent discussion about the
perspectives offered by future SKA radio surveys]{Gitti_2015}.

\vspace{-0.25in}
\section*{Acknowledgments}
\vspace{-0.01in}

We thank Fabrizio Brighenti and the anonymous referee for useful
comments. GB acknowledge support from von Humboldt Foundation and
PRIN--INAF2014.

\vspace{-0.25in}
\bibliographystyle{mn2e}
\bibliography{bibliography}

\begin{thebibliography}{}

\bibitem[\protect\citeauthoryear{{Akritas} \& {Bershady}}{{Akritas} \&
  {Bershady}}{1996}]{Akritas_1996}
{Akritas} M.~G.,  {Bershady} M.~A.,  1996, \apj, 470, 706

\bibitem[\protect\citeauthoryear{{Aleksi{\'c}}, {Alvarez}, {Antonelli} \& {et
  al.}}{{Aleksi{\'c}} et~al.}{2012}]{Aleksic_2012}
{Aleksi{\'c}} J.,  {Alvarez} E.~A.,  {Antonelli} L.~A.,    {et al.} 2012, \aap,
  541, A99

\bibitem[\protect\citeauthoryear{{Arnaud}}{{Arnaud}}{1996}]{Arnaud_1996}
{Arnaud} K.~A.,  1996, in {Jacoby} G.~H.,  {Barnes} J.,  eds, Astronomical Data
  Analysis Software and Systems V Vol.~101 of Astronomical Society of the
  Pacific Conference Series, {XSPEC: The First Ten Years}.
p.~17

\bibitem[\protect\citeauthoryear{{Brunetti} \& {Jones}}{{Brunetti} \&
  {Jones}}{2014}]{Brunetti_2014}
{Brunetti} G.,  {Jones} T.~W.,  2014, International Journal of Modern Physics
  D, 23, 30007

\bibitem[\protect\citeauthoryear{{Brunetti} \& {Lazarian}}{{Brunetti} \&
  {Lazarian}}{2011}]{Brunetti-Lazarian_2011}
{Brunetti} G.,  {Lazarian} A.,  2011, \mnras, 412, 817

\bibitem[\protect\citeauthoryear{{Carilli} \& {Taylor}}{{Carilli} \&
  {Taylor}}{2002}]{Carilli-Taylor_2002}
{Carilli} C.~L.,  {Taylor} G.~B.,  2002, \araa, 40, 319

\bibitem[\protect\citeauthoryear{{Cavagnolo}, {Donahue}, {Voit} \&
  {Sun}}{{Cavagnolo} et~al.}{2009}]{Cavagnolo_2009}
{Cavagnolo} K.~W.,  {Donahue} M.,  {Voit} G.~M.,    {Sun} M.,  2009, \apjs,
  182, 12

\bibitem[\protect\citeauthoryear{{Dickey} \& {Lockman}}{{Dickey} \&
  {Lockman}}{1990}]{Dickey_1990}
{Dickey} J.~M.,  {Lockman} F.~J.,  1990, \araa, 28, 215

\bibitem[\protect\citeauthoryear{{Fabian}}{{Fabian}}{1994}]{Fabian_1994}
{Fabian} A.~C.,  1994, \araa, 32, 277

\bibitem[\protect\citeauthoryear{{Fabian}}{{Fabian}}{2012}]{Fabian_2012}
{Fabian} A.~C.,  2012, \araa, 50, 455

\bibitem[\protect\citeauthoryear{{Feretti}, {Giovannini}, {Govoni} \&
  {Murgia}}{{Feretti} et~al.}{2012}]{Feretti_2012}
{Feretti} L.,  {Giovannini} G.,  {Govoni} F.,    {Murgia} M.,  2012, \aapr, 20,
  54

\bibitem[\protect\citeauthoryear{{Gaensler}, {Landecker}, {Taylor} \& {POSSUM
  Collaboration}}{{Gaensler} et~al.}{2010}]{Gaensler_2010}
{Gaensler} B.~M.,  {Landecker} T.~L.,  {Taylor} A.~R.,    {POSSUM
  Collaboration} 2010, in American Astronomical Society Meeting Abstracts \#215
  Vol.~42 of Bulletin of the American Astronomical Society, {Survey Science
  with ASKAP: Polarization Sky Survey of the Universe's Magnetism (POSSUM)}.
p. \#470.13

\bibitem[\protect\citeauthoryear{{Giacintucci}, {Markevitch}, {Venturi} \& {et
  al.}}{{Giacintucci} et~al.}{2014}]{Giacintucci_2014}
{Giacintucci} S.,  {Markevitch} M.,  {Venturi} T.,    {et al.} 2014, \apj, 781,
  9

\bibitem[\protect\citeauthoryear{{Gitti}, {Brighenti} \& {McNamara}}{{Gitti}
  et~al.}{2012}]{Gitti_2012}
{Gitti} M.,  {Brighenti} F.,    {McNamara} B.~R.,  2012, Advances in Astronomy,
  2012, 6

\bibitem[\protect\citeauthoryear{{Gitti}, {Brunetti}, {Feretti} \&
  {Setti}}{{Gitti} et~al.}{2004}]{Gitti_2004}
{Gitti} M.,  {Brunetti} G.,  {Feretti} L.,    {Setti} G.,  2004, \aap, 417, 1

\bibitem[\protect\citeauthoryear{{Gitti}, {Brunetti} \& {Setti}}{{Gitti}
  et~al.}{2002}]{Gitti_2002}
{Gitti} M.,  {Brunetti} G.,    {Setti} G.,  2002, \aap, 386, 456

\bibitem[\protect\citeauthoryear{{Gitti}, {Tozzi}, {Brunetti}, {Cassano} \& {et
  al.}}{{Gitti} et~al.}{2015}]{Gitti_2015}
{Gitti} M.,  {Tozzi} P.,  {Brunetti} G.,  {Cassano} R.,    {et al.} 2015, in
  Advancing Astrophysics with the Square Kilometre Array (AASKA14) {The SKA
  view of cool-core clusters: evolution of radio mini-halos and AGN feedback}.
p.~76

\bibitem[\protect\citeauthoryear{{Govoni}, {Murgia}, {Markevitch}, {Feretti},
  {Giovannini}, {Taylor} \& {Carretti}}{{Govoni} et~al.}{2009}]{Govoni_2009}
{Govoni} F.,  {Murgia} M.,  {Markevitch} M.,  {Feretti} L.,  {Giovannini} G.,
  {Taylor} G.~B.,    {Carretti} E.,  2009, \aap, 499, 371

\bibitem[\protect\citeauthoryear{{Johnston-Hollitt}, {Govoni}, {Beck} \& {et
  al.}}{{Johnston-Hollitt} et~al.}{2015}]{Johnston_2015}
{Johnston-Hollitt} M.,  {Govoni} F.,  {Beck} R.,    {et al.} 2015, Advancing
  Astrophysics with the Square Kilometre Array (AASKA14), p.~92

\bibitem[\protect\citeauthoryear{{Mazzotta} \& {Giacintucci}}{{Mazzotta} \&
  {Giacintucci}}{2008}]{Mazzotta_2008}
{Mazzotta} P.,  {Giacintucci} S.,  2008, \apjl, 675, L9

\bibitem[\protect\citeauthoryear{{McNamara} \& {Nulsen}}{{McNamara} \&
  {Nulsen}}{2007}]{Mcnamara_2007}
{McNamara} B.~R.,  {Nulsen} P.~E.~J.,  2007, \araa, 45, 117

\bibitem[\protect\citeauthoryear{{McNamara} \& {Nulsen}}{{McNamara} \&
  {Nulsen}}{2012}]{Mcnamara_2012}
{McNamara} B.~R.,  {Nulsen} P.~E.~J.,  2012, New Journal of Physics, 14, 055023

\bibitem[\protect\citeauthoryear{{Myers}}{{Myers}}{2014}]{Myers_2014}
{Myers} S.~T.,  2014, in Exascale Radio Astronomy {The Karl G. Jansky VLA Sky
  Survey (VLASS): A Scientific and Technical Proving Ground for the SKA Era}.
p. 10302

\bibitem[\protect\citeauthoryear{{Peterson} \& {Fabian}}{{Peterson} \&
  {Fabian}}{2006}]{Peterson_2006}
{Peterson} J.~R.,  {Fabian} A.~C.,  2006, \physrep, 427, 1

\bibitem[\protect\citeauthoryear{{Pfrommer} \& {En{\ss}lin}}{{Pfrommer} \&
  {En{\ss}lin}}{2004}]{Pfrommer_2004}
{Pfrommer} C.,  {En{\ss}lin} T.~A.,  2004, \aap, 413, 17

\bibitem[\protect\citeauthoryear{{Rafferty}, {McNamara}, {Nulsen} \&
  {Wise}}{{Rafferty} et~al.}{2006}]{Rafferty_2006}
{Rafferty} D.~A.,  {McNamara} B.~R.,  {Nulsen} P.~E.~J.,    {Wise} M.~W.,
  2006, \apj, 652, 216

\bibitem[\protect\citeauthoryear{{Sutherland} \& {Dopita}}{{Sutherland} \&
  {Dopita}}{1993}]{Sutherland_1993}
{Sutherland} R.~S.,  {Dopita} M.~A.,  1993, \apjs, 88, 253

\bibitem[\protect\citeauthoryear{{van Weeren}, {Intema}, {Lal} \& {et
  al.}}{{van Weeren} et~al.}{2014}]{vanWeeren_2014}
{van Weeren} R.~J.,  {Intema} H.~T.,  {Lal} D.~V.,    {et al.} 2014, \apjl,
  786, L17

\bibitem[\protect\citeauthoryear{{Zandanel}, {Pfrommer} \& {Prada}}{{Zandanel}
  et~al.}{2014}]{Zandanel_2014}
{Zandanel} F.,  {Pfrommer} C.,    {Prada} F.,  2014, \mnras, 438, 124

\bibitem[\protect\citeauthoryear{{Zhuravleva}, {Churazov}, {Schekochihin} \&
  {et al.}}{{Zhuravleva} et~al.}{2014}]{Zhuravleva_2014}
{Zhuravleva} I.,  {Churazov} E.,  {Schekochihin} A.~A.,    {et al.} 2014, \nat,
  515, 85

\bibitem[\protect\citeauthoryear{{ZuHone}, {Markevitch}, {Brunetti} \&
  {Giacintucci}}{{ZuHone} et~al.}{2013}]{Zuhone_2013}
{ZuHone} J.~A.,  {Markevitch} M.,  {Brunetti} G.,    {Giacintucci} S.,  2013,
  \apj, 762, 78

\end{thebibliography}

\end{document}